# Saving Lives: Design and Implementation of Lifeline Emergency Ad Hoc Network

Se-Hang Cheong[a], Yain-Whar Si[1,b], Leong-Hou U[c]

Department of Computer and Information Science, University of Macau
dit.dhc@lostcity-studio.com[a], fstasp@umac.mo[b], ryanlhu@umac.mo[c]

## Abstract

**Purpose** – This paper proposes a system for automatically forming ad hoc networks using mobile phones and battery-powered wireless routers for emergency situations. The system also provides functions to send emergency messages and identify the location of victims based on the network topology information.

**Design/methodology/approach** – OLSR is used to instantly form an ad hoc emergency network based on WiFi signals from mobile phones of the victims, backup battery-powered wireless routers preinstalled in buildings, and mobile devices deployed by search and rescue teams. The proposed system is also designed to recover from partial crash of network and nodes lost.

**Findings** – Experimental results demonstrate the effectiveness of the proposed system in terms of battery life, transmission distance and noises.

**Originality/value** – A novel message routing schedule is proposed for conserving battery life. A novel function to estimate the location of a mobile device which sent an emergency message is proposed in this paper.

**Key Words:** Ad hoc network, natural disaster, emergency response, position locating, Optimized Link State Routing

## 1. INTRODUCTION

Cellular networks constitute the backbones of today telecommunication industry. In some of the developed countries, the percentage of the population owning a mobile phone can be as high as 88% (Poushter, 2016). There are real cases where mobile phones have been contributed for successful rescues in recent catastrophic disasters. Examples of such cases have been reported in history (Dybwad, 2010; Hacaoglu et al., 2011; Napallacan, 2011). However, Public Switched Telephone Networks (PSTN), including cellular networks could be disrupted during natural disasters, such as earthquakes, hurricane, tsunami, etc. In such situations, mobile phone users may not be able make emergency calls since cellular signal is likely to be unavailable due to the destruction of land-based network infrastructures. Although we cannot assume that all people trapped under debris will have access to their mobile phones, it is possible that some of them may still have access to their mobile phones when they are trapped under the debris or when they are waiting to be rescued.

---

[1] Corresponding author: Yain-Whar Si. Present address: Faculty of Science and Technology, University of Macau, E11-4022, Avenida da Universidade, Taipa, Macau, China.

In this paper, we describe the design and implementation of a novel emergency ad hoc network called Lifeline which does not reply on the functioning of PSTN. An ad hoc network is a collection of wireless mobile nodes (or routers) dynamically forming a temporary network without relying on any existing network infrastructure or centralized administration (Toh, 2002). The key motivation of the work presented in this paper is to design and implement an efficient and highly responsive emergency ad hoc network for residential areas, including densely populated places such as shopping malls, schools, hospitals, sport stadiums, and government offices. By using the proposed emergency ad hoc network, we hope that the victims who are trapped under debris should be able to send messages instantly in the aftermath of the natural disasters, such as earthquakes, hurricane, tsunami.

The main function of Lifeline is to instantly form an ad hoc emergency network based on WiFi signals from mobile phones of the victims, backup battery-powered wireless routers preinstalled in buildings, and mobile devices deployed by search and rescue teams. The devices used by search and rescue teams are not only limited to mobile phones. These devices may include laptops, battery-powered routers, and other WiFi enabled portable devices. We use mobile phones as the major components in our network design since they are ubiquitous and increasingly affordable to much of the population. When there is no cellular signal to make an emergency call, victims can use Lifeline Apps pre-installed in his/her mobile phone to connect to the emergency ad hoc networks formed by similar Lifeline installed devices. After that, he/she can send the emergency messages to an emergency station via the ad hoc network formed by the Lifeline. Emergency station can be any WiFi enabled device/computer operated by a rescuer. In addition, Lifeline Apps can capture and send on-site photos to assist search and rescue teams. Furthermore, by using the last known positions of backup battery-powered wireless routers, we also propose algorithms for locating the approximate positions of the mobile devices (or victims) in an ad hoc network. Lifeline also has the ability to recover from partial crash of network and nodes lost. To recover from partial crash, battery-powered wireless routers in Lifeline are designed to backup messages and automatically boot up when they encounter the network failure.

Mobile devices have only limited battery life and the golden time of 72 hours is extremely important for survivors as well as for the members of search and rescue teams during disasters. Therefore, conserving the battery life of mobile devices within the disaster areas is crucial in emergency situations. The consumption of battery power can be high when nodes in an ad hoc network are designed to forward the messages to nearby devices. Therefore, a power saving message routing schedule in which inner nodes can be turn off occasionally into sleep or energy saving mode to reduce the messages is proposed in this paper. By using the message routing schedule, Lifeline can prolong the battery power of boundary nodes and maximize their throughput to store or forward messages from the inner nodes.

We have developed a prototype system of Lifeline and performed extensive experiments to evaluate its efficiency. We have evaluated performance of message forwarding, Backup Message Module, and battery consumption in our experiments. These experiments are performed for different scenarios including routers and mobile phones.

The summary of the technical contributions of the paper are as follows:
1. The design and implementation of a novel emergency ad hoc network called Lifeline is proposed in this paper.
2. Lifeline is able to constructs a self-organizing emergency ad hoc network based on WiFi when cellular signal is unavailable.

3. A novel message routing schedule for conserving battery life is proposed in this paper.
4. Lifeline is able to recover from partial crash of network and nodes lost.
5. A novel function to estimate the location of a mobile device which sent an emergency message is proposed in this paper.
6. The effectiveness of the proposed system is tested by measuring the battery consumption and number of emergency messages forward per milliseconds.

This paper is the extension version of the conference paper (Cheong et al., 2011). We review the related work in section 2. The design and implementation of Lifeline is discussed in section 3. In Section 4, experiments are conducted to evaluate the performance of the Lifeline. Finally, we summarize our ideas in section 5.

## 2. RELATED WORK

Emergency ad hoc network have been studied in different research fields (Fujiwara et al., 2005; Khan et al., 2009; Mao et al., 2011; Mocito et al., 2010; Nilsson et al., 2010; Ramrekha et al., 2012). The emergency ad hoc networks are designed to operate even if the infrastructure facilities of cellular system is damaged or corrupted. The objectives of the emergency ad hoc networks are to provide connectivity and to copes with communication congestion even in disaster situations.

Pawelczak et al. (2005) proposed a dynamic bandwidth assignment per channel in wireless emergency communication because each channel is assigned a fixed bandwidth in traditional radio frequency partitioning. Their approach is useful in emergency cases in which lots of distress messages are waiting for transmission. When emergency cases occur, their approach can increase the bandwidth of the channel which most likely transferring emergency messages dynamically. Therefore, the transfer speed and the throughput of the channel can be increase instantly in emergency cases. Moreover, Durresi et al. (2005) proposed an emergency protocol for inter-vehicle communication. Sensors are installed in vehicles to gather information in real time. Their proposed system broadcasts emergency messages to the network when a distress signal is raised. A number of other emergency systems were proposed for vehicles in recent years (Abboud et al., 2009; Hartenstein et al., 2001; Resta et al., 2007; Sahoo et al., 2009). In addition, Malan et al. (2004) proposed an ad network for a medical care application. The system can be deployed on wearable devices with vital sign sensor installed. These sensors can track patient's status and location, while simultaneously operating as active tags.

Bulusu et al. (2000) proposed a positioning system based on a uniform grid of short-range radio frequency (RF) transceiver. In their system, a random node in the network is able to localize itself by estimating its distance to the well-known positions of the closest radio receivers. Capkun et al. (2001) proposed a relative coordinate system for an ad hoc network that does not rely on Global Positioning System (GPS). The algorithm uses the distances between the nodes to build a relative coordinate system. Their distances are retrieved from the Time of Arrival (TOA) and Signal Strength data of the ad hoc network. Niculescu et al. (2003) proposed a positioning algorithm in ad hoc networks using distance vector (DV) based methods namely DV-hop and DV-distance methods. The distances of DV-hop are measured using the hop count. The DV-distances between nodes are measured using radio signal strength instead of the hop count of two nodes.

There are various routing protocols developed for ad hoc networks such as OLSR (Benzaid et al., 2002; "Optimized Link State Routing Protocol (OLSR)," 2003),

BATMEN (Neumann et al.), and AODV (Perkins et al., 2003). According to recent reports (Barolli et al., 2009; Kulla et al., 2010), the performance of OLSR is stable than other protocols. OLSR (Optimized Link State Routing) protocol is an optimization of the pure link state algorithm. It is one of the most popular link state algorithms among open source protocols ("Optimized Link State Routing Protocol (OLSR)," 2003). OLSR is highly extensible since it can work well with many extensions, such as link quality and fisheye-algorithm (Pei et al., 2000). OLSR protocol is a pro-active routing protocol that constructs a route for data transmission by storing a routing table inside every node of the network. OLSR is also able to integrate localization services ("Optimized Link State Routing Protocol (OLSR)," 2003; Soga et al., 2010) thus it can be used to estimate the location of nodes in the network.

In this paper, OLSR (Optimized Link State Routing) protocol is used to form the Lifeline emergency ad hoc network. Every node in the OLSR constructs a topology control (TC) packet for their neighbors list in turn to update the routing table stored in the node. The TC packet is used to determine whether the neighbor is alive or dead. A node in OLSR protocol is considered as a neighbor if and only if it can be reached via a bi-directional link. Moreover, OLSR protocol does not broadcast control packets to the entire network. Elected nodes (relays) are the only nodes allowed in OLSR to flood messages to neighbor nodes. This relay mechanism can significantly reduce the total amount of control traffic and improve the power consumption of OLSR networks (Verbree et al., 2010).

## 3. SYSTEM OVERVIEW

Lifeline can be installed in mobiles phones, battery-powered wireless routers, and computers. It also supports various scenarios for forwarding messages from a sender to the emergency station. The system design of Lifeline is depicted in Figure 1. Emergency station is a WiFi enabled computer with Lifeline backend system installed. Lifeline Apps is designed for mobile phones with Android operating system. An embedded version of Lifeline (Embedded Lifeline) is also developed for routers which support OpenWrt ("OpenWrt,").

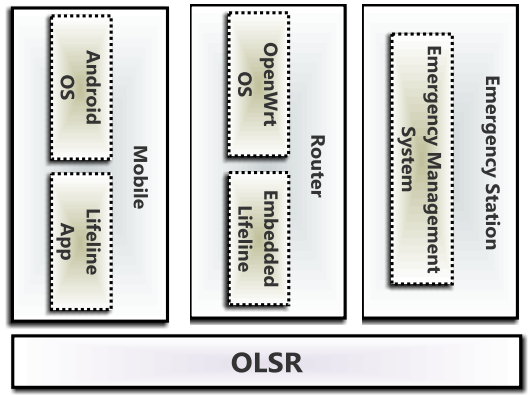

Figure 1 System design

### 3.1 Lifeline App

Lifeline App was developed for mobile phones with Android operating system. Features implemented in Lifeline App include:

1. Forming emergency ad networks if no cellular signal is available: mobile phones can be configured to form a new emergency ad hoc network or join an available one.
2. Displays the total time the user has been waiting for help since the last emergency message is sent out.
3. Allows sending and receiving of emergency messages through the emergency ad hoc networks. The messages will be forwarded to available emergency stations nearby, see Figure 2(a).
4. The user can take a photo with the mobile phone and send it with the emergency message. Although the exact location may not be known, the situation of the surrounding environment of the trapped victim may provide some hints to the rescue team, see Figure 2(b).
5. Personal information can be configured in advance when Lifeline App is installed. When an emergency is sent out, pre-composed personal information will be included, see Figure 2(b).

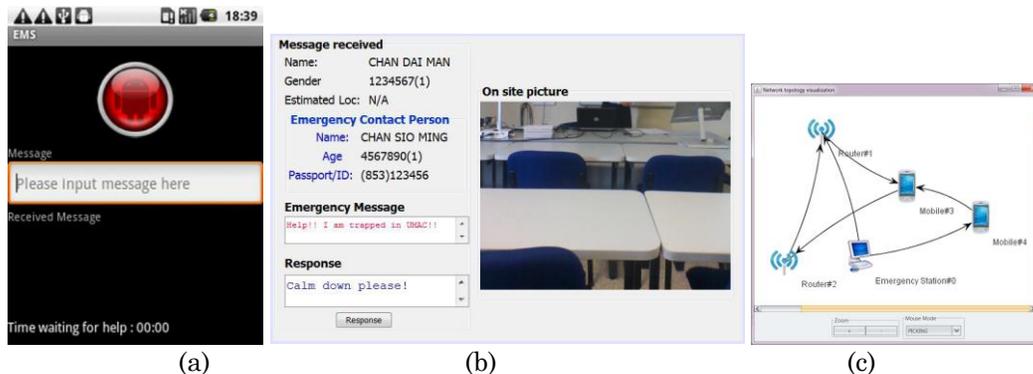

Figure 2 (a) The screenshot of emergency message editor, (b) Reply function in emergency station, (c) The real-time network topology viewer.

### 3.2 Embedded version of Lifeline application (Embedded Lifeline)

Regardless of the brands, mobile phones have limited coverage of WiFi signal and battery life. Therefore, mobile phones cannot be expected to function for days when WiFi function is turned on. This situation can be more serious especially when a large number of messages are being routed within the ad hoc network. In contrast, wireless routers have wider WiFi signal range compared to mobile phones and can be paired with larger (backup) batteries. Therefore, we developed an embedded version of Lifeline application for WiFi enabled battery-powered wireless routers. In addition to wider signal range, wireless routers have also other significant advantages. They are inexpensive compared to dedicated mobile emergency cellular network stations. These wireless routers are also portable and rescue teams can quickly deploy them when an emergency case occurred. Lifeline App in mobile phones allows the transmission of messages to the emergency station across heterogeneous devices including mobile phones and wireless routers. Likewise, an embedded version of Lifeline is also developed for routers based on an open source firmware called OpenWrt ("OpenWrt,"). We use OLSR ("Optimized Link State Routing Protocol (OLSR)," 2003) as the routing protocol in the firmware. Features implemented in Embedded Lifeline application for battery-powered wireless routers include:

1. Battery-powered wireless routers can be configured to boot up as an emergency node when electrical supply is interrupted.

2. Allows forwarding of emergency messages from the victims to the emergency stations.
3. Battery-powered wireless routers can be configured with rules to backup emergency messages received.
4. Battery-powered wireless routers can be preinstalled in residential areas and they can be configured with a physical location. When an ad hoc network is formed, the location estimation protocol can collect this information for approximating the position of the message sender.
5. When the battery-powered wireless router has low battery, it will pass the remaining messages to nearby devices and reject incoming messages gradually through the ACPI (Advanced Configuration and Power Interface) of the routers.

Wireless routers with backup battery can be mounted in residential units or public buildings in earthquakes prone areas. For example, these battery-powered wireless routers can be installed along side with emergency exit lights in hospitals, schools, or public places. When earthquake strikes and electricity is cut off, the battery-powered wireless routers will be reset after a certain amount of time. After that, these battery-powered wireless routers will automatically boot up to form emergency hoc networks with other nearby devices containing Lifeline programs. Apart from using this system in disaster areas, normal buildings can also be installed such battery-powered wireless routers to allow forming of ad hoc networks during fire or severe storms.

### 3.3 Emergency station

Emergency station is a WiFi enabled computer with Lifeline backend system installed. Features implemented in Lifeline system for emergency station include:
1. Allows rescuers to receive and reply messages instantly.
2. Rescuers can monitor the topology of the nodes in the emergency ad hoc network, see Figure 2(c).
3. Lifeline provides functions for the rescuers to approximate the location of the message sender if the nearest location of the battery-powered wireless router is known.

### 3.4 Lifeline ad hoc network

Figure 3 illustrates the possible combination of routes for forwarding messages. Messages in Lifeline can be transmitted from the mobile phones and will be routed to an emergency station. In scenario 1 at the center of Figure 3, we illustrate the case in which emergency messages are routed by mobile phones only. The scenario 2 and scenario 3 depict the situations where emergency messages are routed through a series of mixed combination of mobile phones and battery-powered wireless routers. The scenario 4 illustrates the situation in which emergency messages are routed by battery-powered wireless routers.

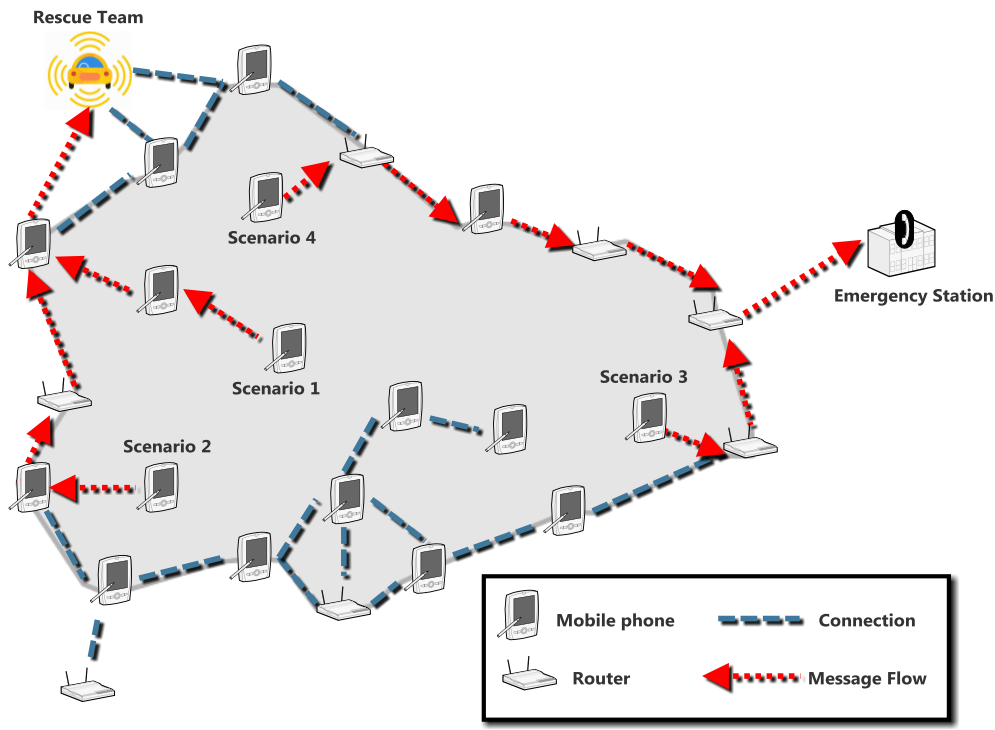

Figure 3 Scenarios of forwarding messages to the emergency station.

We describe the details of Lifeline ad hoc network in the following sections. Section 3.4.1 describes the implementation details of how emergency messages are forwarded in the Lifeline App. Section 3.4.2 describes the technical details of Forward Message Module and Priority Queue. These components are implemented in battery-powered wireless routers to prioritize and forward emergency messages. Section 3.4.3 describes the strategies implemented in Backup Message Module for storing emergency messages when emergency stations are unreachable due to overloading or network crash. Section 3.4.4 describes the emergency boot strategies implemented in the battery-powered wireless routers to enable them to boot up and form emergency ad hoc networks when the disaster strikes. Section 3.4.5 describes the Position Locating Algorithm to estimate the location of the victims based on the messages received.

### 3.4.1 Sending emergency messages with Lifeline App in mobile phones

Lifeline App in victims' mobile phones provides features for sending emergency messages. In the implementation of Lifeline, a specified range of IP addresses are reserved for emergency stations only. These addresses will not be used by the Lifeline Apps and battery-powered wireless routers. Once the Lifeline App is initiated, it sends Topology Control (TC) packets to discover link state information of the emergency ad hoc network. When the Lifeline App is connected to the emergency ad hoc network, an IP address is assigned to the device so that the App can send emergency messages through OLSR protocol. Emergency messages are forwarded hop-by-hop until they reach the emergency station. The reply messages from the emergency station to each victim's device is transported in the same way. The message forwarding is based on the routing path formed by nodes selected as Multipoints Relays (MPRs) ("Optimized Link State Routing Protocol (OLSR)," 2003).

### 3.4.2 Forward Message Module with Priority Queue in battery-powered wireless routers

Forward Message Module illustrated in Figure 4(a) was implemented in the Embedded Lifeline program for routers. Inside the module, Receive Module accepts TCP packets (by a TCP server listening to the port 33333) and filter emergency messages. The router will only pass emergency messages to the Schedule Module and other messages will be ignored. The TCP server was implemented in ANSI C programming language and the emergency messages were formatted based on XML. Schedule Module determines the priority of emergency messages by analyzing the message header and emergency messages will be passed into Forward Module. To increase the performance of message dispatching, we designed five levels of priority (starting from 0 to 4) in Lifeline and each level of priority has an independent queue as illustrated in Figure 4(b). Forward Module sends out messages which are previously stored in the priority queues. If the destination is unreachable, the message will be transferred back to the Schedule Module and the priority of the message will be decreased by one.

Due to the limited memory available in routers, there is insufficient space to maintain all messages in the Random Access Memory (RAM). From our experiments, we find that only a few megabytes of RAM are available for use by the Lifeline program in routers. Therefore, the lower the priority, the higher the chance emergency messages in the priority queues will be swapped out to the secondary storage of the routers. That is, if an emergency message is in the low priority queues such as in Priority-3 and Priority-4 queues, the message will be swapped out to the secondary storage. On the other hand, messages in the lower priority queue of Schedule Module will be gradually moved into higher priority queues when messages are being sent out. That is, all messages in Priority-3 queue are moved to Priority-2 queue, messages in Priority-4 queue are moved to Priority-3 queue, and so on. When there is no emergency message left in Priority-0 and Priority-1 queues, any messages which are previously swapped out to the secondary storage will be swapped in and sent to the destination again.

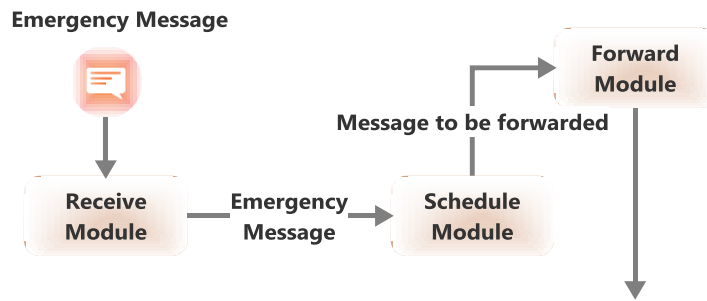

(a)

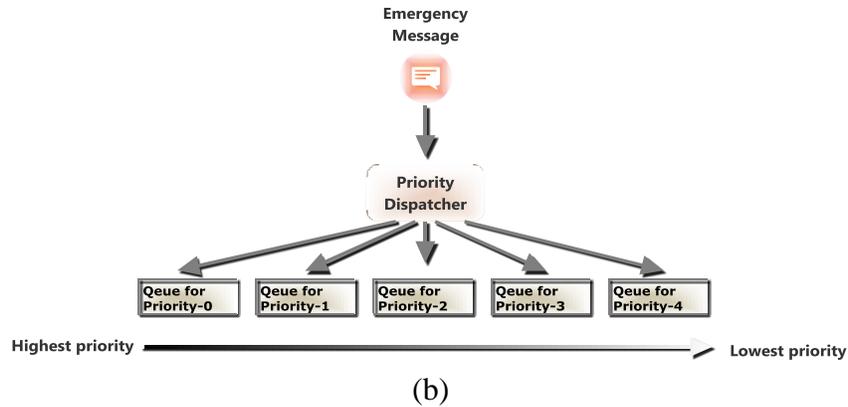

(b)

Figure 4 (a) Overview of Forward Message Module, (b) Priority queues of emergency messages.

### 3.4.3 Backup Message Module in battery-powered wireless routers

The emergency station could be unreachable when the network is overloaded or partial network crash occurs. When the emergency station is unreachable or the battery of the router is low, all emergency messages will be stored in the permanent storage of the router. To achieve this objective, Backup Message Module is implemented in battery-powered wireless routers. The options for Backup Message Module are listed in Table 1.

Table 1 Options for Backup Message Module

| Option number | Options | Priority |
|---|---|---|
| 1 | Always backup the received message. | 1 |
| 2 | Always backup the message after it is forwarded. | 1 |
| 3 | Backup the message if battery is less than *%. 0<*<=100 | 2 |
| 4 | Backup the message with priority higher than *%. 0<=*<=4 | 2 |
| 5 | Backup the message if the load of the current device is higher than *%. 0<*<100. | 3 |
| 6 | Backup the message if the load on the source (sender) is higher than *%. 0<*<100. | 3 |

For example, if a battery-powered wireless router is enabled with two options "always backup message after forwarded" and "backup message if the load of the current device is higher than 5%", and the current load of system is 10%, when an emergency message was received, the message will be stored as a backup because the first option has a higher priority than the second option. The low priority options will be discarded. However, OLSR is a designed for a decentralized network. In addition, we can only retrieve the neighborhood information through OLSR protocol and the global view of the entire network is unavailable. Therefore, the message sender will attach its load information in its emergency message.

### 3.4.4 Emergency Boot Module in battery-powered wireless routers

Three emergency boot strategies are implemented in the battery-powered wireless routers so that these routers can boot up and form emergency ad hoc networks after the disaster.

*Strategy 1*: Battery-powered wireless routers initialize the process to form the emergency ad hoc network when AC power supply is interrupted or the backup battery starts power supply to the router. This strategy depends on the availability of detecting the source of power supply. However, accidental booting can occur when the supply of AC power is unstable or the AC power is switched off by someone.

*Strategy 2*: Battery-powered wireless routers initialize the process to form an emergency ad hoc network when the level of backup battery drops. This can be done by monitoring the backup battery level. The power consumption of backup battery shouldn't change dramatically when the AC power is in use. If a dramatic change in the backup battery consumption is detected, then we can conclude that the AC power has been disrupted. This strategy requires the exact consumption data of the battery and may trigger false alarm if the quality of backup battery is poor or the battery is old.

*Strategy 3*: Battery-powered wireless routers initialize the process to form an emergency ad hoc network by scanning nearby peers. These routers scan its neighbors constantly and tries to determine whether there are emergency stations nearby or devices that are in emergency mode (i.e. when the Lifeline program on the device is initiated). When it identifies one, the router initializes the process to form an emergency ad hoc network. One of the main advantages of this strategy is that it allows routers to provide general functions in normal situation. When a disaster occurs, these routers can form the emergency ad hoc network automatically. Figure 5 illustrates an example of scanning nearby nodes and forming the emergency ad hoc network.

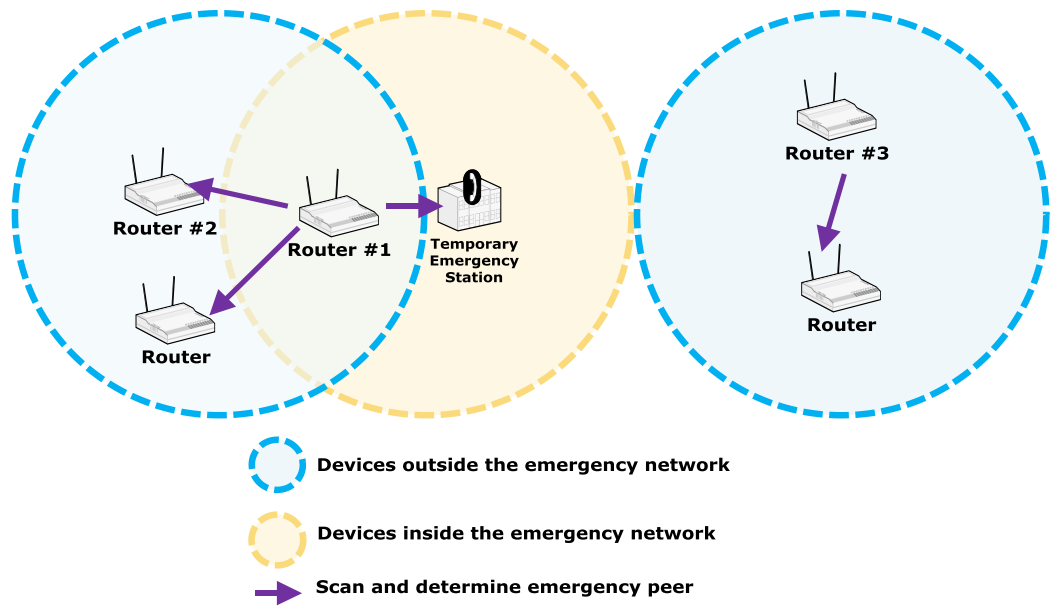

Figure 5 Scanning nearly peers for forming an emergency ad hoc network.

Two scenarios are illustrated in the Figure 5. The first scenario illustrates the router which connect to the temporary emergency station and the second scenario does not. In the first scenario, Router#1 scans nearby WiFi devices and analyzes all information (i.e. SSID (Service Set Identifier), signal strength, signature, etc.) retrieved from these devices. First, Router#1 checks the signature of every device retrieved since a device in emergency mode is assigned with a distinct signature. Next, Router#1 will check whether the signature of emergency mode in devices is detected. In this scenario, Router#1 will *switch* to the emergency mode (i.e. temporary emergency station detected) so that it can establish to the emergency ad hoc network. Router#1 will then update its signature so that indicating Router#1 has become a node in the emergency mode. Moreover, Router#2 will perform similar procedures as Router#1 did. That is, Router#2 checks the signature of every device retrieved. Router#2 will then *join* to the emergency ad hoc network because Router#2 detected Router#1 is a node of the emergency mode.

In the second scenario, Router#3 scans nearby WiFi devices and analyzes all information retrieved from these devices. Router#3 cannot detect any temporary emergency station or the node in the emergency mode. Therefore, Router#3 waits for a predetermined interval and repeat scanning of the nearby devices again.

### 3.4.5 Position Locating

Position Locating Algorithm allows rescue teams and emergency stations to estimate the location of the victims who have sent emergency messages. In these situations, we assume that GPS signal is not available since the victims could be trapped under collapsed buildings. The estimated positions of the battery-powered routers are available in Lifeline since battery-powered wireless routers can be configured with known locations. Once an emergency network is formed, physical locations can be queried from those routers.

Passive and active approaches of transmitting position locating message are illustrated in Figure 6. The example of passive approach is illustrated in Router B's network. When a mobile phone queries the physical location, it broadcasts a "WHERE AM I?" to devices within N-Hops. If there is a router pre-configured with known physical location is found within N-Hops, the router will reply back to the mobile phone with a message containing its position. If no such device is found, the mobile phone cannot determine the physical location up to N-Hops. The example of active approach is illustrated in Router A's network. In this approach, if a router is configured with known physical location, then the router sends its physical location to newly joined mobile phones within N-Hops. This operation will be executed every time whenever the topology of the emergency ad hoc network is changed.

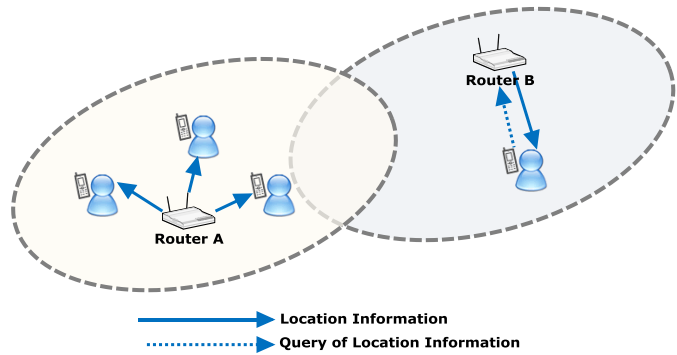

Figure 6 Message transmission for position locating.

## 4. EXPERIMENTS

We evaluated the performance and the battery power consumption of Lifeline ad hoc network in terms of the number of routers and mobile phones, size and number of messages. The Embedded Lifeline program are installed on ASUS WL500gPv2 routers in our prototype. The Lifeline App for mobile phones are developed for Huawei U8150 IDEOS mobile phone and the program is compatible with Android SDK API version larger than 18. Screenshots of the mobile phone and routers given in Figure 7.

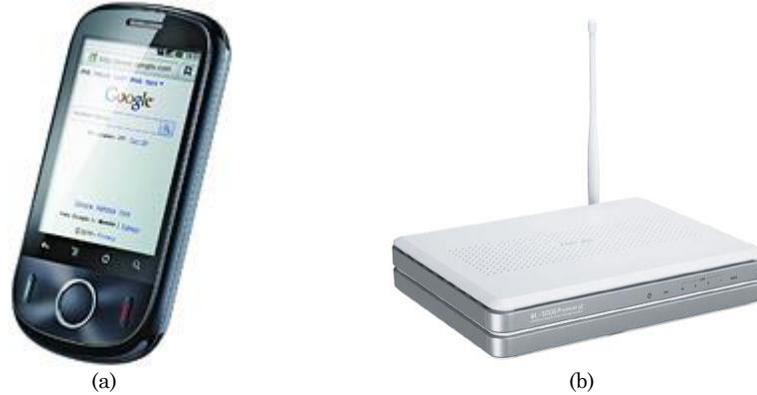

(a)  (b)
Figure 7 (a) Screenshot of Huawei U8150 IDEOS mobile phone (b) Screenshot of ASUS WL500gPv2

### 4.1 Performance of message forwarding

In our experiments, the size of an emergency message is set to 255 bytes. 1000 and 10000 emergency messages for each priority are generated and forwarded on the ad hoc network. We measure the average running time of each priority in milliseconds. The objective of this experiment is to evaluate the performance with varying composition of devices with respect to the size of emergency messages and the priority of emergency messages.

#### 4.1.1 Performance of forwarding emergency messages in an ad hoc network containing a single router (Set-up A)

In this experiment, we use an ASUS WL500gPv2 router to send emergency messages to the router itself by using a loopback address. This experiment is to evaluate the performance when the messages are forward in a router with a loopback network interface.

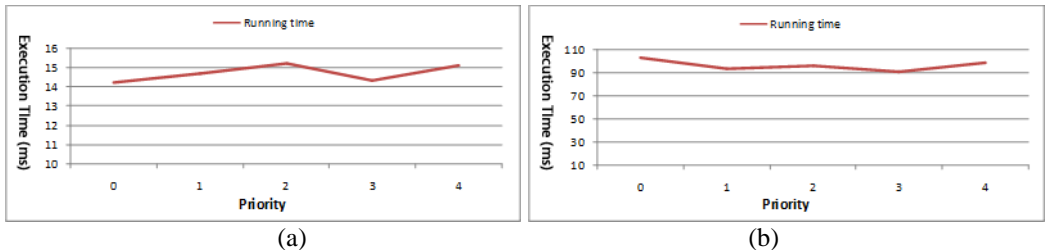

(a)  (b)
Figure 8 (a) sending 255-byte emergency messages 1000 times for each priority, (b) sending 255-byte emergency messages 10000 times for each priority.

Figure 8 illustrates the results of forwarding 1000 and 10000 emergency messages per priority. From these results, we can observe that the execution times are similar for different priorities. The average execution time for 1000 emergency messages is about 15ms (milliseconds) and the average execution time for 10000 emergency messages is about 95ms.

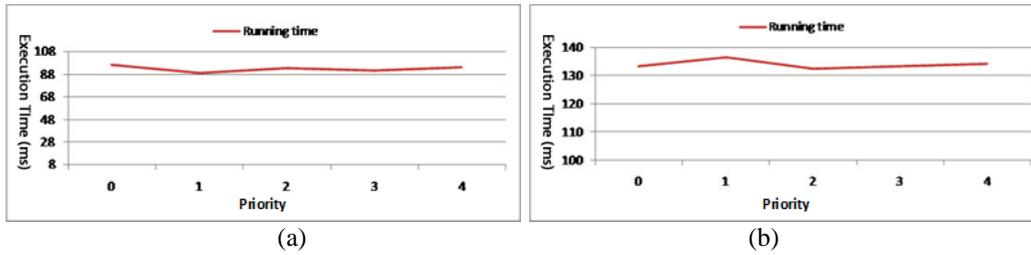

(a) (b)

Figure 9 (a) sending random size emergency messages 1000 times for each priority, (b) sending random size emergency messages 10000 times for each priority.

Figure 9 illustrates the results of another experiment where the settings are the same as the previous one except that the size of emergency messages are randomly generated from the range of 10 bytes to 255 bytes. Form the results, we can observe that the running time is higher than experiments illustrated in Figure 8. The average execution times for 1000 and 10000 emergency messages are approximately 88ms and 133ms.

We also evaluated other combinations of settings in a router with respect to the size of emergency messages and priorities. These experiment results are shown in Table 2. The experimental result shows that the execution time is slow when we send emergency messages with random priority. The experimental result also shows that the execution time is faster when the size of emergency messages is constant.

Table 2 Experiment results of forward emergency messages in a router.

| Size | Number of messages | Priority | Average execution time (ms) |
|---|---|---|---|
| Constant size (255 bytes) | 1000 | Fixed priority (Priority-0) | 29.6 |
| | | Random priority | 33.5 |
| | 10000 | Fixed priority (Priority-0) | 196.2 |
| | | Random priority | 208.7 |
| Random size (randomly generated from 10 bytes to 255 bytes) | 1000 | Fixed priority (Priority-0) | 28.4 |
| | | Random priority | 30.2 |
| | 10000 | Fixed priority (Priority-0) | 188.6 |
| | | Random priority | 193.5 |

### 4.1.2 Performance of forwarding emergency messages in an ad hoc network containing four routers (Set-up B)

In this experiment, we used four routers to construct an emergency ad hoc network. Routers are placed in the same room, the distance among them is less than 5 meters. This objective of this experiment is to evaluate the performance when the messages are forwarded in four routers with short distance among these routers.

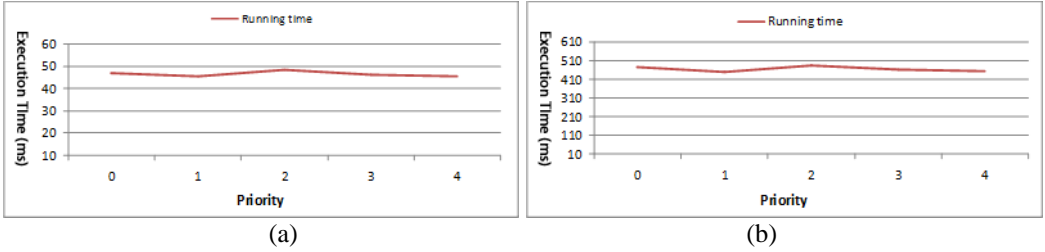

(a) (b)

Figure 10 (a) sending 255-byte emergency messages 1000 times for each priority, (b) sending 255-byte emergency messages 10000 times for each priority.

Figure 10 illustrates the results of forwarding 1000 and 10000 emergency messages per priority. From these results, we find that the execution time for different priorities are also similar. The average execution time for 1000 emergency messages is about 45ms and it is 4 times slower than the experimental result of a single router. The average execution time for 10000 emergency messages is about 45ms and it is about 5 times slower than the experimental result of a single router.

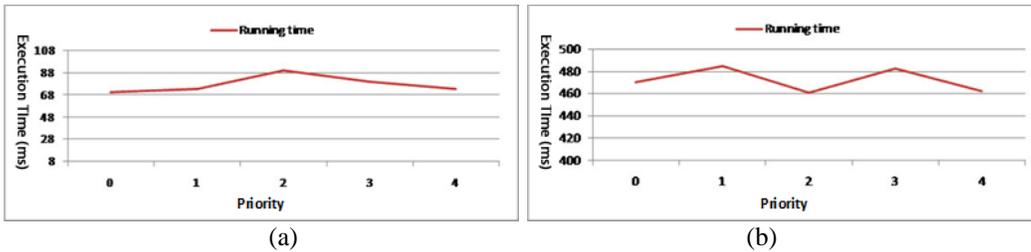

(a) (b)

Figure 11 (a) sending random size emergency messages 1000 times for each priority, (b) sending random size emergency messages 10000 times for each priority.

Figure 11 illustrates the results of another experiment. The experiment settings are same as the previous one except that the emergency message sizes are randomly generated from the range of 10 bytes to 255 bytes. Form the experiment results, we can observe that the running time is higher than the experiments illustrated in Figure 10. The average execution time for 1000 emergency messages is about 70ms and the average execution time for 10000 emergency messages is about 460ms.

We also evaluated other combinations of experiment settings with respect to varying messages sizes and priorities. These experiment results are shown in Table 3.

Table 3 Experimental result of forwarding messages in an ad hoc network containing four routers.

| Size | Number of messages | Priority | Average running time (ms) |
|---|---|---|---|
| Constant size (255 bytes) | 1000 | Fixed priority (Priority-0) | 80.7 |
| | | Random priority | 93.5 |
| | 10000 | Fixed priority (Priority-0) | 567.6 |
| | | Random priority | 703.5 |
| Random size (randomly generated from 10 bytes to 255 | 1000 | Fixed priority (Priority-0) | 69.8 |
| | | Random priority | 80.4 |
| | 10000 | Fixed priority | 1304.2 |
| bytes) | | (Priority-0) | |
| | | Random priority | 1658.5 |

### 4.1.3 Performance of forwarding emergency messages in an ad hoc network containing four routers and two mobile phones (Set-up C)

In this experiment, four routers and two mobile phones are used to construct the emergency ad hoc network. Routers and two mobile phones are placed in the same room and the distance among them is less than 5m.

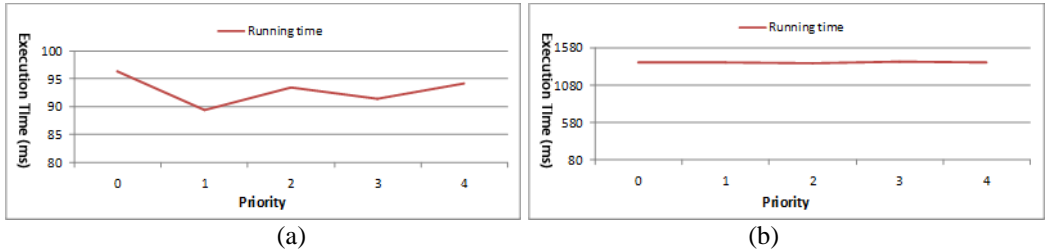

(a) (b)

Figure 12 (a) sending 255-byte emergency messages 1000 times for each priority, (b) sending 255-byte emergency messages 10000 times for each priority.

Figure 12 illustrates the results of forwarding 1000 and 10000 emergency messages per priority. From the experiment results, we can observe that the execution time of forwarding messages for all priorities are similar. The average execution time for 1000 emergency messages is about 95ms and it is about 2 times slower than the experimental result of four routers. The average execution time for 10000 emergency messages is about 1380ms and it is about 3 times slower than result of four routers.

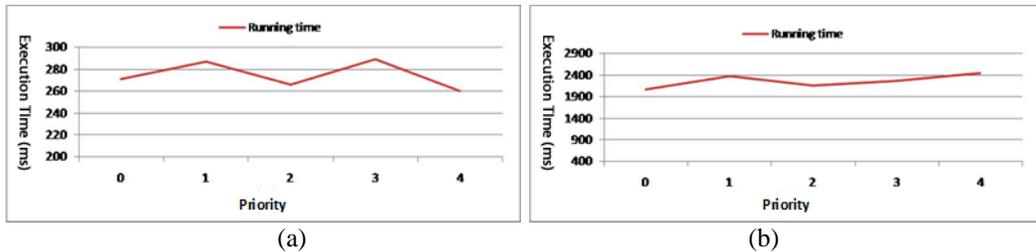

(a) (b)

Figure 13 (a) sending random size of emergency messages 1000 times for each priority, (b) sending random size of emergency messages 10000 times for each priority.

Figure 13 illustrates the results of another experiment. The experiment settings are same as the previous one except that the message size is random generated between 10 bytes and 255 bytes. The running time is higher than experiments illustrated in Figure 12. The average execution time for 1000 emergency messages is about 270ms and it is 4 times slower than the experimental result of four routers. The average execution time for 10000 emergency messages is about 2200ms and it is about 5 times slower than the experimental result of four routers.

We also evaluated other combinations of experiment settings with an ad hoc network containing four routers and two mobile phones. These experiment results are shown in Table 4.

Table 4 Experiment results of forwarding messages in an ad hoc network containing four routers and two routers.

| Size | Number of | Priority | Average running |
|---|---|---|---|

| | messages | | time (ms) |
|---|---|---|---|
| Constant size (255 bytes) | 1000 | Fixed priority (Priority-0) | 357.2 |
| | | Random priority | 375.1 |
| | 10000 | Fixed priority (Priority-0) | 2071.0 |
| | | Random priority | 2263.4 |
| Random size (randomly generated from 10 bytes to 255 bytes) | 1000 | Fixed priority (Priority-0) | 459.1 |
| | | Random priority | 541.9 |
| | 10000 | Fixed priority (Priority-0) | 2498.0 |
| | | Random priority | 2694.7 |

### 4.1.4 Performance of forwarding emergency messages in an ad hoc network containing four routers and two mobile phones with long transmission distances (Set-up D)

In this experiment, we used four routers and two mobile phones to construct the emergency ad hoc network. Four routers and two mobile phones are placed 60m apart each other. The transmission errors captured from the experiment is shown in Table 5. The purpose of this experiment is to evaluate the performance network for long transmission distances.

Table 5 The transmission errors of messages.

| Test | Number of messages | Error in sending | Error in receiving |
|---|---|---|---|
| **Constant size of emergency message and the priority** | 1000 | 4 | 2 |
| | 10000 | 18 | 32 |
| **Random emergency message size and same priority** | 1000 | 3 | 1 |
| | 10000 | 17 | 34 |

We also evaluated other combinations of experiment involving four routers and two mobile phones with respect to the size of emergency messages and priorities. These experiment results are shown in Table 6.

Table 6 Experiment results of forwarding emergency messages in an ad hoc network containing four routers and two routers with long transmission distance.

| Size | Number of messages | Priority | Average running time (ms) |
|---|---|---|---|
| Constant size (255 bytes) | 1000 | Fixed priority (Priority-0) | 686.0 |
| | | Random priority | 705.3 |
| | 10000 | Fixed priority (Priority-0) | 7351.1 |
| | | Random priority | 7931.8 |
| Random size (randomly generated from | 1000 | Fixed priority (Priority-0) | 861.2 |
| | | Random priority | 1035.1 |

| | | Fixed priority (Priority-0) | 7301.9 |
| 10 bytes to 255 bytes) | 10000 | Random priority | 7505.8 |

**4.2 Performance of Backup Message Module**

The emergency station could be unreachable when the network is overloaded or partial network crash occurs. When the emergency station is unreachable or the battery of the router is low, all emergency messages will be stored in the permanent storage of the router. To achieve this objective, Backup Message Module is implemented in battery-powered wireless routers. Our experiments evaluate the execution time when different options of Backup Message Module are used. We test two options for the backup message module and the procedures for the experiments are as follows:
1. Routers are configured with option-1 as the default option of Backup Message Module. That is, all routers will store all emergency messages received. We then record the back-up time of 1000 and 10000 emergency messages.
2. Routers are configured with option-4 using value 0 as the default option of Backup Message Module. That is, routers will store all messages with priority 0. We then record the back-up time of 1000 and 10000 emergency messages.

**4.2.1 Performance of Backup Message Module in a single Router (Set-up E)**

In this experiment, we use an ASUS WL500gPv2 router to test the performance of Backup Message Module. The experimental results are summarized in Table 7. According to the results, a router takes about 50ms to store 1000 messages for option-1 and takes about 32ms to log down 1000 messages for option-4. When we increase the number of emergency messages from 1000 to 10000, the execution time is 400.7ms and 270.1ms respectively which is about 8 times slower than the execution time for 1000 emergency messages.

Table 7 The experimental results of backup emergency module in a single router.

| Size | Number of messages | Option | Average running time (ms) |
|---|---|---|---|
| Constant size (255 bytes) | 1000 (200 messages are Priority-0) | 1 | 50.2 |
| | | 4 with value = 0 | 32.0 |
| | 10000 (2000 messages are Priority-0) | 1 | 400.7 |
| | | 4 with value = 0 | 270.1 |

**4.2.2 Performance of Backup Message Module in an ad hoc network containing four routers (Set-up F)**

Four routers are used to form an emergency ad hoc network in this experiment. Routers are placed in the same room and the distance among them is less than 5 meters. The experiment settings are described in section 4.2. The experimental results are summarized in the Table 8. When we compare the experimental results to the section 4.2.1, we find that the performance of four routers are 7 times slower than that of a single router for the configuration option-1 (in both 1000 and 10000 emergency messages). Performance of four routers is also about 7 times slower than that of a single router for the configuration option-4 (in both 1000 and 10000 emergency messages).

Table 8 The experimental results of backup emergency module for four routers.

| Size | Number of messages | Option | Average Running time (ms) |
|---|---|---|---|
| Constant size (255 bytes) | 1000 (200 messages are Priority-0) | 1 | 1003.5 |
| | | 4 with value = 0 | 504.4 |
| | 10000 (2000 messages are Priority-0) | 1 | 3045.6 |
| | | 4 with value = 0 | 1230.2 |

### 4.2.3 Backup Message Module in an ad hoc network containing four routers and two mobile phones (Set-up G)

Four routers and two mobile phones are used to construct an emergency ad hoc network in this experiment. Routers and mobile phones are placed in the same room and the distance among them is less than 5m. The experimental settings are described in section 4.2. The experimental results are summarized in Table 9. When we compare the results to the section 4.2.2, we find that current results are about 8 times slower than that of four routers for the configuration option-1 (in both 1000 and 10000 emergency messages). The experimental results also show that it is about 7 times slower than that of four routers for the configuration option-4 (in both 1000 and 10000 emergency messages).

All the experimental results from section 4.2.1, 4.2.2 and 4.2.3 show that the execution time of option-1 is higher than option-4.

Table 9 The experimental results of backup emergency module for an ad hoc network containing four routers and two mobile phones.

| Size | Number of messages | Option | Average Running time (ms) |
|---|---|---|---|
| Constant size (255 bytes) | 1000 (200 messages are Priority-0) | 1 | 7561.7 |
| | | 4 with value = 0 | 4204.2 |
| | 10000 (2000 messages are Priority-0) | 1 | 15807.4 |
| | | 4 with value = 0 | 9726.3 |

### 4.3 Battery Testing

We used four routers, two mobile phones and a PC acting the temporary emergency station to form an emergency ad hoc network in this experiment. These devices are placed in the same room, the distance among them is less than 5 meters. Huawei Ideos Mobile phones and ASUS WL500gPv2 routers are used for the experiments. The objective of this experiment is to evaluate the battery consumption with respect to the interval of emergency messages and the status of the mobile phones.

The first experiment was to test the battery life of a mobile phone in an idle situation. In this experiment, we turned on the mobile phone and initialized the

Lifeline application. After that the screen of the mobile phone was dimmed to test the battery life. We then recorded the elapsed time until the battery was completely flat. According to our experiment, we found that the mobile phone can last for about 15 hours without any emergency messages being forwarded (i.e. there is no workload on the phone). When we turned on the screen of the mobile phone and repeated the above experiment, it lasted for about 7 hours.

We also tested the maximum battery life with different workloads in our experiments. In this experiment, we used a laptop to connect to the emergency ad hoc network. We then loaded the laptop with a Lifeline App so that it can act as a mobile phone. Next, we use the laptop to generate emergency messages repeatedly and the messages are forwarded to the emergency station via a series of mobile phones. Each message is 255 bytes long and they are forwarded with an interval of 10 seconds, 1 minute, and 5 minutes. To extend the battery life, we dimmed the screen of mobile phones during this experiment. The messages are forwarded via the mobile phones to the emergency station until the mobile phones have no more battery power. The experimental results are illustrated in Figure 14. From the results, we can observe that mobile phones can last for about 7, 11, and 13 hours for forwarding in 10s, 60s, and 300s interval.

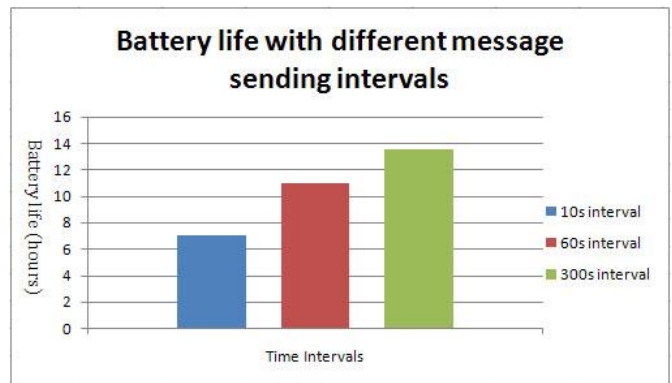
Figure 14 Battery life tests for different intervals.

5. **CONCLUSION**

In this paper, we describe the design and implementation of a novel system called Lifeline which is capable of forwarding and receiving emergency messages via an emergency ad hoc network. The system provides disaster victims with an alternative way to seek help when cellular signals are unavailable during emergency situations. Lifeline can automatically form an emergency ad hoc network when cellular networks are unavailable during or aftermath of disasters. Lifeline is able recover from partial crash of network and nodes lost. Lifeline also provides functions to estimate the approximation location of a sender (potential victim) where an emergency message was originated.

To evaluate the performance of message forwarding, we set up four scenarios; a router (Set-up A), four routers (Set-up B) and four routers and two mobile phones (Set-up C). For each scenario, various size of emergency messages per priority are then forwarded in the experiment. Furthermore, we also tested the effect of long transmission distances in which four routers and two mobile phones (Set-up D) are placed 60 meters apart each other. According to our experimental results, Set-up D is 3 times slower than Set-up C in which devices are placed 5 meters apart and Set-up

C is 5 times slower than Set- up B. We also observed that Set-up B is 5 times slower than the Set-up A.

In addition, we evaluated the performance of Backup Message Module for four scenarios; a router (Set-up E), four routers (Set-up F) and four routers and two mobile phones (Set-up G). According to the experimental results, Set-up G is approximately 8 times slower than Set-up F and Set-up E is about 4.5 times faster than Set-up F.

Moreover, we also evaluated the battery consumption test. The mobile phone in our experiment can last for about 15 hours without any workload while the screen is dimed. If we turn on the screen of the mobile phone and repeat above experiment, it can last for about 7 hours. Furthermore, the mobile phone can last for about 7, 11 and 13 hours when they are forwarded within an interval of 10 seconds, 60 seconds and 300 seconds respectively and each message size is set to 255 bytes.

A potential direction for future work is to minimize the energy consumption during the message forwarding in OLSR. Since Lifeline is designed for emergency cases, the battery life of the devices is crucial. The longer the battery can last, the higher the chance the user could be rescued. Therefore, further improvements can be done by reducing the overhead of message forwarding to increase the battery life of the mobile phones. Another direction is to develop a more precise algorithm for locating the victims by integrating other approaches.

**ACKNOWLEDGEMENT**

This research was funded by the Research Committee of University of Macau, grant MYRG2016-00148-FST and MYRG2017-00029-FST.